\documentclass[preprint,superscriptaddress]{revtex4-2} 
\usepackage{graphicx} 
\usepackage{amsmath, cases} 
\usepackage[version=3]{mhchem} 
\usepackage{longtable} 

\usepackage{mathtools} 
\usepackage{url} 
\usepackage{verbatim} 
 
\usepackage{epstopdf} 
\usepackage{ upgreek }

\usepackage{hyperref} 
\usepackage{cleveref} 

\usepackage[normalem]{ulem} 
\usepackage{xcolor}

\begin{document} 

\title{Electronic Properties of Ladder-Type Phenylenes Studied by Thermally-Assisted-Occupation Density Functional Theory} 

\author{Tsu-Hsuan Wang} 
\affiliation{Department of Physics, National Taiwan University, Taipei 10617, Taiwan} 

\author{Jeng-Da Chai} 
\email[Author to whom correspondence should be addressed. Electronic mail: ]{jdchai@phys.ntu.edu.tw} 
\affiliation{Department of Physics, National Taiwan University, Taipei 10617, Taiwan} 
\affiliation{Center for Theoretical Physics and Center for Quantum Science and Engineering, National Taiwan University, Taipei 10617, Taiwan} 
\affiliation{Physics Division, National Center for Theoretical Sciences, Taipei 10617, Taiwan} 

\date{\today}

\begin{abstract} 

Accurate prediction of the electronic properties of extended $\pi$-conjugated systems with radical character remains a major theoretical challenge. To overcome this challenge, thermally-assisted-occupation density 
functional theory (TAO-DFT) has been developed in recent years. In this study, we employ TAO-DFT to systematically explore the electronic properties of ladder-type phenylenes ($n$-LPs), containing $n$ linearly 
fused four-membered carbon rings. Our calculations reveal that all investigated $n$-LPs (with $n$ = 1–40) possess singlet ground states. With increasing $n$-LP length, we observe a transition from the nonradical 
character of shorter $n$-LPs to the polyradical character of longer $n$-LPs. Notably, we identify a distinct damped periodic oscillation in the radical character of $n$-LPs, featuring highly correlated hotspots at specific 
lengths ($n = 3k - 1$, where $k$ are positive integers). The real-space representation of active orbitals confirms this global electronic behavior is governed by the intrinsic structural periodicity and completely 
delocalized nature of the active orbitals along the central ladder framework. These findings provide fundamental insights into the size-dependent electronic properties of $n$-LPs, guiding their rational design for future 
nanomaterial applications. 

\end{abstract} 

\maketitle 

\clearpage

\section{Introduction} 

Carbon is a highly versatile element, capable of forming diverse structures through $sp$, $sp^2$, and $sp^3$ hybridization. This bonding flexibility has facilitated the successive discovery of various carbon 
nanoallotropes \citep{ref-carbon1,ref-carbon2}, progressing from zero-dimensional (0D) fullerenes \citep{ref-carbon3,ref-carbon4} and one-dimensional (1D) carbon nanotubes \citep{ref-carbon5,ref-carbon6} to 
two-dimensional (2D) graphene \citep{ref-carbon7,ref-carbon8,ref-carbon9}. Among them, graphene, a single planar sheet of carbon atoms arranged in a honeycomb lattice with $sp^2$ hybridization, has attracted 
intense research interest. Its delocalized $\pi$-electron network yields exceptional electronic, thermal, and mechanical properties, offering significant potential for next-generation nanoelectronics and 
spintronics \citep{ref-graphene1,ref-graphene2}. However, because pristine 2D graphene lacks an intrinsic band gap, its application in logic switching devices, such as field-effect transistors, is severely 
limited \citep{ref-transistors}. Consequently, confining graphene's planar dimensions to the nanoscale, such as graphene nanoribbons (GNRs) \citep{ref-GNR1,ref-GNR2,ref-TAO2}, graphene nano-parallelograms 
(GNPs) \citep{ref-GNP1,ref-GNP2,ref-TAO_GNP}, and graphene quantum rings (GQRs) \citep{ref-GQR1,ref-GQR2,ref-TAO4}, has emerged as an effective strategy for band-gap engineering through quantum 
confinement, finite-size effects, and edge effects \citep{ref-nanosystems1,ref-nanosystems2}. 

While conventional 1D carbonaceous systems, such as GNRs and linear acenes, typically exhibit a monotonic trend in their electronic properties towards a narrow-gap or polyradical bulk 
limit \citep{ref-TAO2,ref-TAO_GNP}, the recent on-surface synthesis of the 2D biphenylene network has directed research attention toward non-benzenoid carbon allotropes \citep{ref-BP1}. Specifically, Fan {\it et al.} 
successfully synthesized an ultraflat biphenylene network via an on-surface interpolymer dehydrofluorination reaction, explicitly demonstrating its intrinsic metallic character, and thereby resolving a long-standing 
theoretical debate \citep{ref-BP1}. Their measurements further revealed that the band gaps of these ribbons rapidly approaches zero with increasing width, an exceptionally steep scaling behavior fundamentally driven 
by the incorporated antiaromatic four-membered rings \citep{ref-BP1}. 

\begin{figure} 
\includegraphics[scale=1.2]{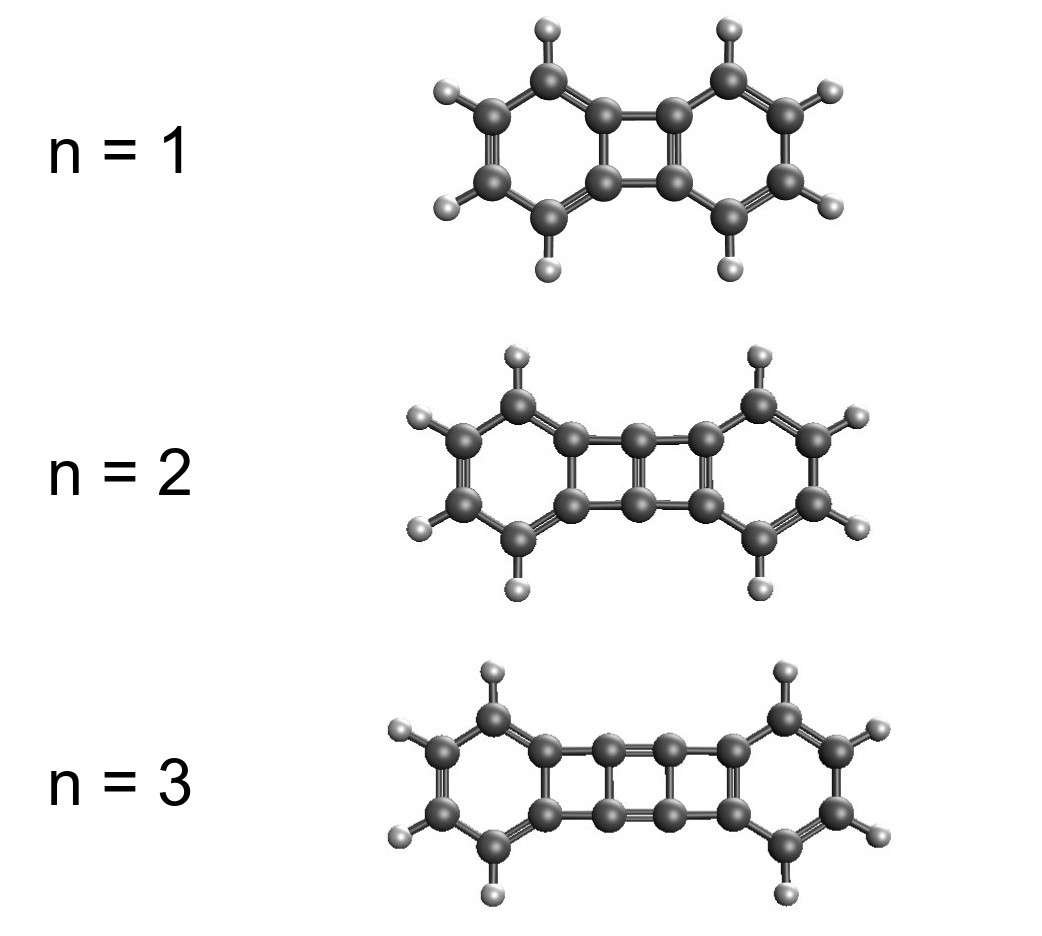} 
\caption{\label{geometry} 
Geometry of $n$-LP (with $n$ = 1–3), containing $n$ linearly fused four-membered carbon rings.} 
\end{figure}

Motivated by these experimental advances, in this work, we investigate the electronic properties of ladder-type phenylenes (LPs) as a prototypical 1D quantum limit of such non-benzenoid networks \citep{ref-BP2}. By 
isolating a consecutive fused cyclobutadiene core, and capping it with stable terminal phenyl groups, we construct a well-defined 1D quantum well. This specific architecture allows us to systematically investigate how 
cumulative antiaromaticity and topological electron frustration dictate the radical character, differing fundamentally from the monotonic boundary conditions of traditional GNRs. As shown in \Cref{geometry}, $n$-LP 
(C$_{10+2n}$H$_{8}$), which contains $N = 68 + 12n$ electrons, exhibits the $D_{2h}$ point-group symmetry. The chain length, denoted by the number ($n$) of linearly fused four-membered carbon rings, directly 
modulates the delocalization of the $\pi$-electron system. The inherent topological frustration and high antiaromatic strain make $n$-LPs excellent models for investigating resonance matching and quantum 
confinement in low-dimensional extended networks \citep{ref-BP1,ref-LP1}. 

However, it remains a significant theoretical challenge to accurately predict the ground-state properties of elongated $\pi$-conjugated systems with radical character (e.g., longer $n$-LPs, as will be shown in this work). 
As the chain length ($n$) increases, the static correlation effects within the extended networks can become pronounced, leading to the emergence of substantial radical character. The unique electronic features of these 
strongly correlated polyradical systems make them highly sought-after candidates for next-generation nanoscale technologies, including organic photovoltaics \citep{ref-PV1,ref-PV2}, molecular 
spintronics \citep{ref-Spintronics}, and nonlinear optics \citep{ref-NLO}. This creates a well-known dilemma in computational methodology. On one hand, conventional single-reference (SR) electronic structure methods 
are generally inadequate for describing these strongly correlated states. For instance, Kohn-Sham density functional theory (KS-DFT) \citep{ref-KS1,ref-KS2} is computationally efficient, and widely used in conjunction 
with standard exchange-correlation (xc) functionals, such as the local density approximation (LDA), generalized gradient approximation (GGA), and hybrid xc functionals. However, spin-restricted KS-DFT adopting these 
standard xc functionals is inherently incapable of capturing static correlation effects \citep{ref-KS3}. Although applying spin-unrestricted KS-DFT can artificially recover a portion of the static correlation energy through 
spatial symmetry lowering, this approach can suffer from unphysical spin-symmetry breaking and severe spin contamination, when applied to multi-reference (MR) systems (e.g., electronic systems with radical character), 
yielding unreliable predictions \citep{ref-SC1,ref-SC2,ref-Rivero,ref-KS3}. For example, recent calculations on hexagonal GQRs have explicitly demonstrated that spin-unrestricted KS-LDA (i.e., KS-DFT with the LDA xc 
functional) can produce highly spin-contaminated lowest singlet and triplet states as the system size expands \citep{ref-TAO4}. On the other hand, while high-level {\it ab initio} MR electronic structure methods, such as 
the density matrix renormalization group (DMRG) algorithms \citep{ref-SC3,ref-ab2}, variational two-electron reduced-density-matrix (2-RDM) methods \citep{ref-2RDM1,ref-2RDM2}, and complete-active-space (CAS) 
approaches \citep{ref-CAS}, can accurately capture these static correlation effects, their computational cost scales prohibitively with the size of the active space, making them computationally intractable for nanosystems. 

To address this dilemma, thermally-assisted-occupation density functional theory (TAO-DFT) \citep{ref-TAO1} has been developed as an efficient electronic structure method for studying large MR systems. Unlike 
finite-temperature density functional theory (also called Mermin-Kohn-Sham (MKS) theory) \citep{ref-FTDFT,ref-KS2}, TAO-DFT is designed to explore the ground-state properties of large MR systems at the absolute 
zero. By introducing fractional orbital occupation numbers governed by the Fermi-Dirac distribution with a fictitious temperature $\theta$, TAO-DFT explicitly incorporates the static correlation energy via an entropy 
contribution \citep{ref-TAO1}. Besides, the standard LDA \cite{ref-TAO1}, GGA \cite{ref-GGA}, hybrid \citep{ref-hybrid1,Xuan2019,tao-b97-d4} exchange-correlation-$\theta$ (xc$\theta$) functionals (i.e., the combined xc 
and $\theta$-dependent functionals) can be used in TAO-DFT. The fictitious temperature $\theta$ in TAO-DFT can be assigned a fixed value for all electronic systems \cite{ref-new_theta}, or can be computed 
self-consistently for any electronic system \cite{etheta-scf}. 

In recent years, the TAO-DFT framework has been continuously expanded to include various capabilities, such as {\it ab initio} molecular dynamics \citep{ref-TAO_AIMD}, polarizable continuum 
models \citep{ref-TAO_PCM}, real-time extensions \citep{ref-RT_TAO}, and finite-temperature extensions \cite{ft-tao-dft}. Importantly, when the fictitious temperature is set to zero ($\theta = 0$), TAO-DFT naturally 
reduces to the standard KS-DFT framework. Besides, the TAO-DFT orbital occupation numbers (TOONs) closely approximate the natural orbital occupation numbers (NOONs) \citep{ref-TAO1,ref-TAO_PAH,ref-NOON}, 
providing a direct and quantitative measure of the radical character within an electronic system. Crucially, because it captures static correlation reasonably within a spin-restricted formalism, TAO-DFT avoids the unphysical 
spin-symmetry breaking and severe spin contamination issues inherent to conventional unrestricted KS-DFT methods \citep{tao-symmetry}. Furthermore, TAO-DFT maintains a computational complexity comparable to that 
of standard KS-DFT, and does not require a pre-defined active space, making it advantageous for investigating large polyradical networks, such as linear acenes \citep{ref-TAO_acene}, alternant polycyclic aromatic 
hydrocarbons \citep{ref-TAO_PAH}, graphene-based nanosystems \citep{ref-TAO2,ref-TAO3,ref-TAO4,ref-TAO_GNP}, and many 
others \citep{Manassir2020,tao-vib1,HansonHeine2020a,HansonHeine2020b,Bettinger2021,Bettinger2021a,tao-vib2,Lischka2023,Somani2024,tao-vib3,Somani2025,Somani2026}. 

Owing to its computational efficiency and accuracy for large MR systems, in this work, we apply TAO-DFT to systematically investigate the electronic properties of $n$-LPs ($n$ = 1–40). We compute the singlet-triplet 
energy gaps, vertical ionization potentials, vertical electron affinities, fundamental gaps, symmetrized von Neumann entropy, and the real-space representation of the active orbitals. We intend to understand how the 
electronic properties change with increasing $n$-LP length.

\section{Computational Details} 

All electronic property calculations and geometry optimizations are performed using a development version of \textsf{Q-Chem 6.1} \citep{ref-qchem} with the 6-31G(d) basis set. The electronic and structural properties 
of $n$-LPs ($n$ = 1–40) are computed using TAO-LDA (i.e., TAO-DFT with the LDA xc$\theta$ functional) \citep{ref-TAO1}, with a nonempirical fictitious temperature $\theta$ = 9.55301 mhartree (see Table 1 of 
Ref.\ \citep{ref-new_theta}). All orbital visualizations are generated using the \textsf{IQmol} software \citep{ref-IQmol}.

\section{Results and Discussion} 

\subsection{Singlet-Triplet Energy Gap} 

For a neutral molecule, the singlet-triplet energy gap (ST gap), which is the energy difference between the lowest singlet and triplet states, is a key measure for developing photonic and magnetic nanoscale 
devices \citep{ref-ST_device1,ref-ST_device2}. For a molecule with a vanishingly small ST gap, it requires little energy to break the pairing of electrons, implying the radical character. Here, the ST gap ($E_{\text{ST}}$) 
of $n$-LP is computed using 
\begin{equation} 
E_{\text{ST}} = E_{\text{UT}} - E_{\text{US}}, 
\end{equation} 
where $E_{\text{US}}$ and $E_{\text{UT}}$ are the spin-unrestricted TAO-LDA energies of the lowest singlet and triplet states, respectively, of $n$-LP, evaluated at the fully optimized geometries. 

\begin{figure} 
\includegraphics[scale=1.0]{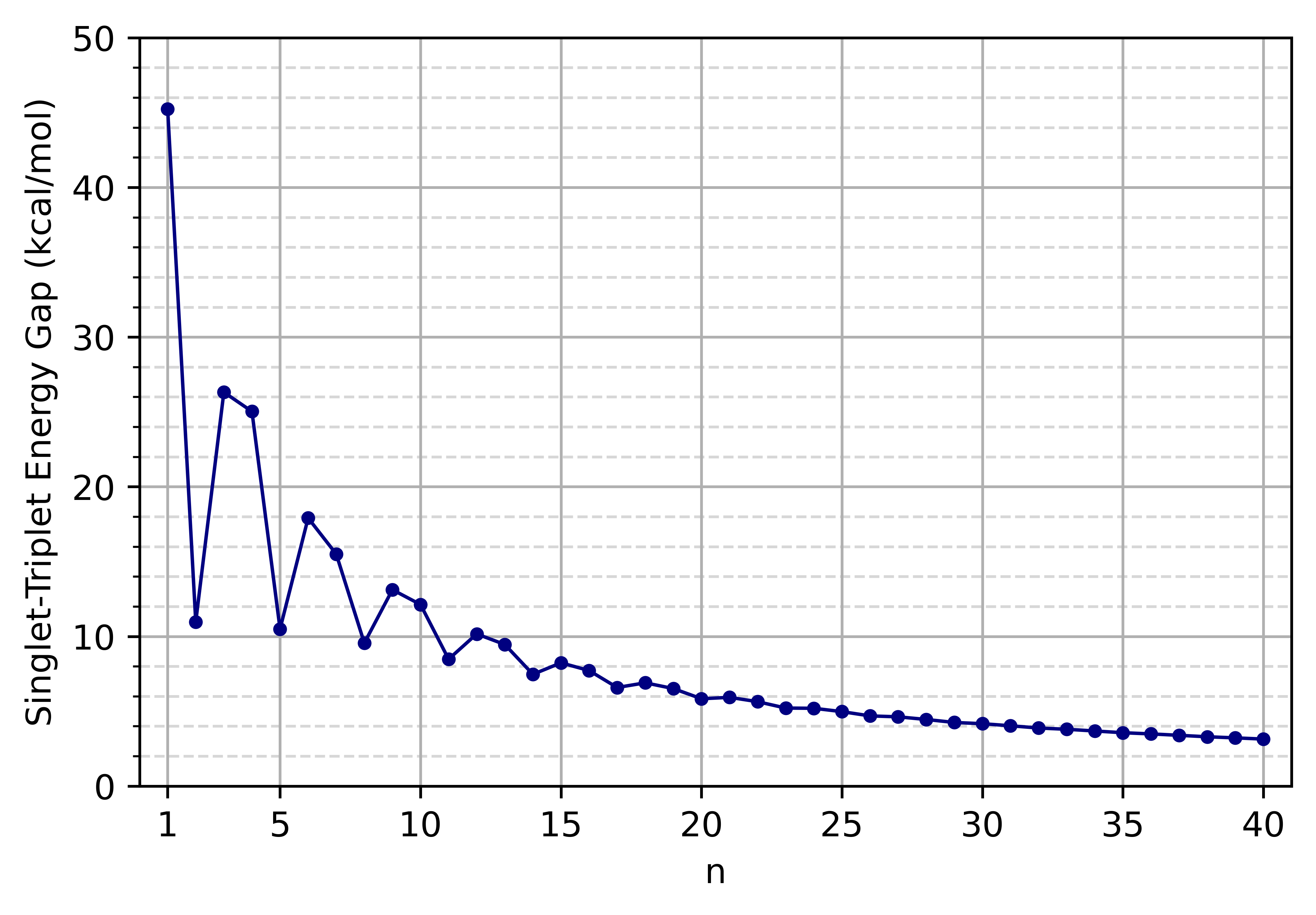} 
\caption{\label{ut-us_tao} 
Singlet-triplet energy gap of $n$-LP, calculated using spin-unrestricted TAO-LDA.} 
\end{figure}

As shown in \Cref{ut-us_tao}, $n$-LPs ($n$ = 1--40) all possess singlet ground states. Although the ST gap generally decreases with increasing chain length, it exhibits a distinct damped oscillatory pattern rather than 
a monotonic decay \citep{ref-TAO2,ref-TAO_GNP} (also see Table S1 in the Supplementary Information (SI)). In particular, a clear periodicity of 3 is observed in the ST gaps of shorter $n$-LPs, with local minima 
appearing at $n = 3k - 1$, where $k$ are positive integers. This observed periodic behavior stems from the interplay between strong static correlation and structural resonance within the two-leg ladder lattice. At critical 
lengths of $n = 3k - 1$, the cumulative electron frustration promotes the unpairing $\pi$-electrons, maximizing the topological frustration and leading to strong static correlation across the fully delocalized 
$\pi$-framework with very small ST gaps. Conversely, at lengths satisfying $n = 3k$, the system undergoes a substantial gap recovery. This periodic restoration of closed-shell stability can be rationalized by 
constructive quantum interference along the multiple pathways of the ladder framework \citep{ref-ST1}, coupled with the optimal periodic matching of Clar's aromatic sextets \citep{ref-ST2}. At these geometric 
resonance points, the wavefunctions effectively overcome the antiaromatic confinement, stabilizing a completely closed-shell configuration and suppressing the radical character. As the length $n$ increases further, 
these oscillations gradually attenuate, and the ST gap approaches zero. This overall reduction of the ST gap in longer $n$-LP implies a transition from a nonradical (closed-shell) character to a pronounced polyradical 
character (characterized by near-degeneracy between electronic states) \citep{ref-TAO2,ref-TAO3,ref-TAO4}. This suggests that the radical character of $n$-LPs can be tuned by modulating the number of 
four-membered rings, facilitating the rational design of molecules with targeted spin properties. 

Conventional SR electronic structure methods, such as spin-unrestricted KS-LDA, frequently suffer from unphysical spin-symmetry breaking when applied to the lowest singlet states of molecules with radical 
character \citep{ref-SC1,ref-SC2,ref-Rivero}. This spatial symmetry lowering produces spin-contaminated solutions, yielding unreliable energy predictions \citep{ref-TAO1,ref-TAO2,ref-TAO4,tao-symmetry}. To confirm 
the absence of artificial spin-symmetry breaking in our spin-unrestricted TAO-LDA calculations, we also compute the lowest singlet energies of $n$-LPs using spin-restricted TAO-LDA, and find that spin-symmetry can 
be retained in our spin-unrestricted TAO-LDA calculations for all the $n$-LPs (with $n$ = 1--40) investigated (i.e., for each $n$-LP, the difference between the spin-restricted and spin-unrestricted TAO-LDA energies is 
within the numerical precision of our calculations). This demonstrates that TAO-LDA provides a physically accurate and reliable description of these systems without the spin contamination artifacts inherent to 
conventional SR electronic structure methods.

\subsection{Vertical Ionization Potential, Vertical Electron Affinity, and Fundamental Gap} 

The fundamental gap is an important electronic property of a neutral molecule, as it directly relates to a molecule's stability, reactivity, and performance in electronic and optoelectronic applications. By definition, 
the fundamental gap ($E_{g}$) is the difference between the vertical ionization potential ($\text{IP}_{v}$) and the vertical electron affinity ($\text{EA}_{v}$): 
\begin{equation} 
E_{g} = \text{IP}_{v} - \text{EA}_{v}. 
\end{equation} 
Physically, $\text{IP}_{v}$ represents the energy required to remove an electron from the neutral molecule, while $\text{EA}_{v}$ denotes the energy released when an electron is added to the neutral molecule. 
At the optimized geometry of ground-state $n$-LP (containing $N$ electrons), we compute the vertical ionization potential as: 
\begin{equation} 
\text{IP}_{v} = E_{N-1} - E_{N}, 
\end{equation} 
and the vertical electron affinity as: 
\begin{equation} 
\text{EA}_{v} = E_{N} - E_{N+1}, 
\end{equation} 
where $E_{N}$ is the total energy of the $N$-electron molecule. These quantities ($\text{IP}_{v}$, $\text{EA}_{v}$, and $E_{g}$) are all evaluated using multiple energy-difference calculations ($\Delta$SCF 
approach) \citep{ref-Deleuze2003,ref-Hajgato2008} with spin-unrestricted TAO-LDA. 

\begin{figure} 
\includegraphics[scale=1.0]{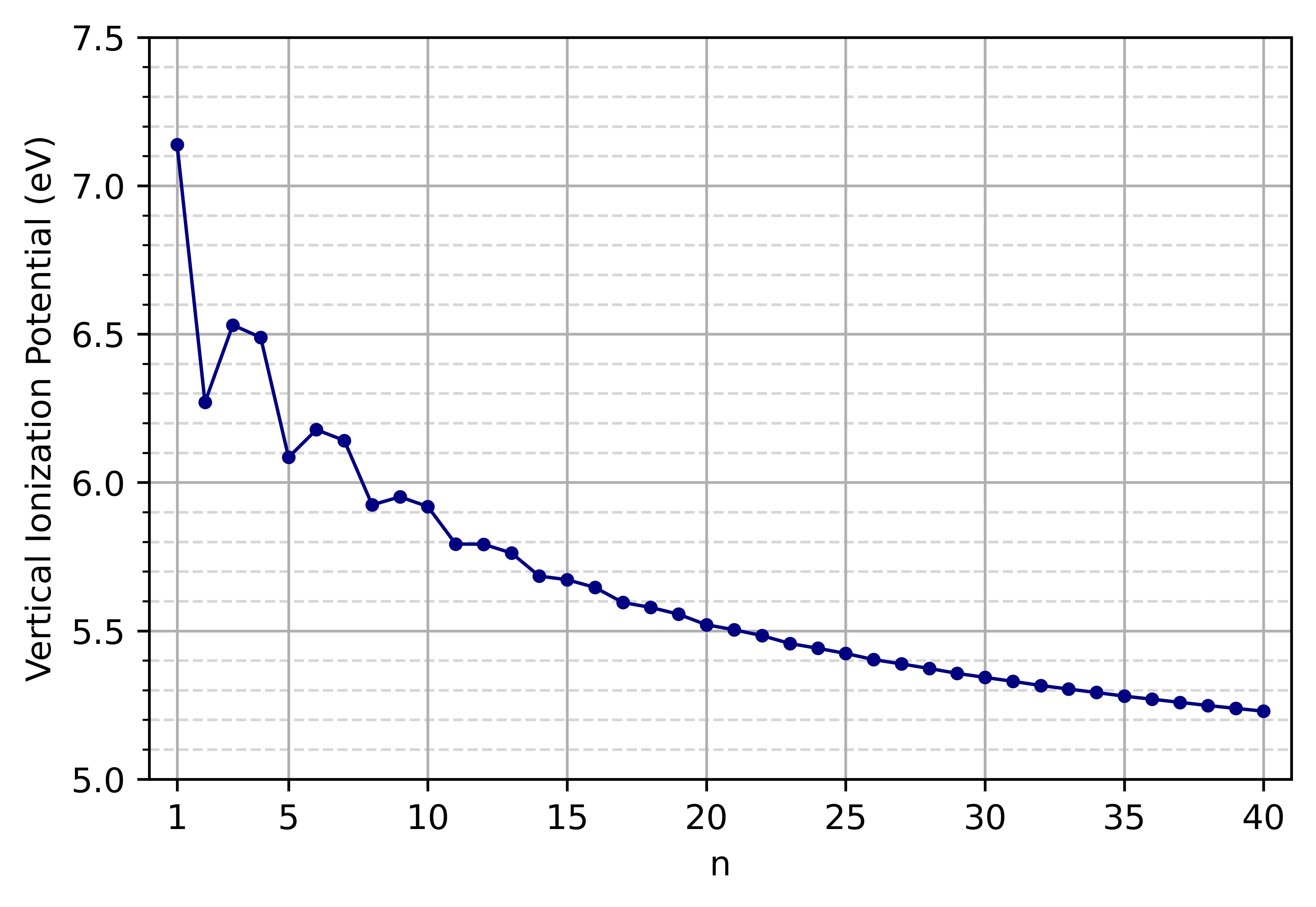} 
\caption{\label{ip} 
Vertical ionization potential of ground-state $n$-LP, calculated using spin-unrestricted TAO-LDA.} 
\end{figure}

\begin{figure} 
\includegraphics[scale=1.0]{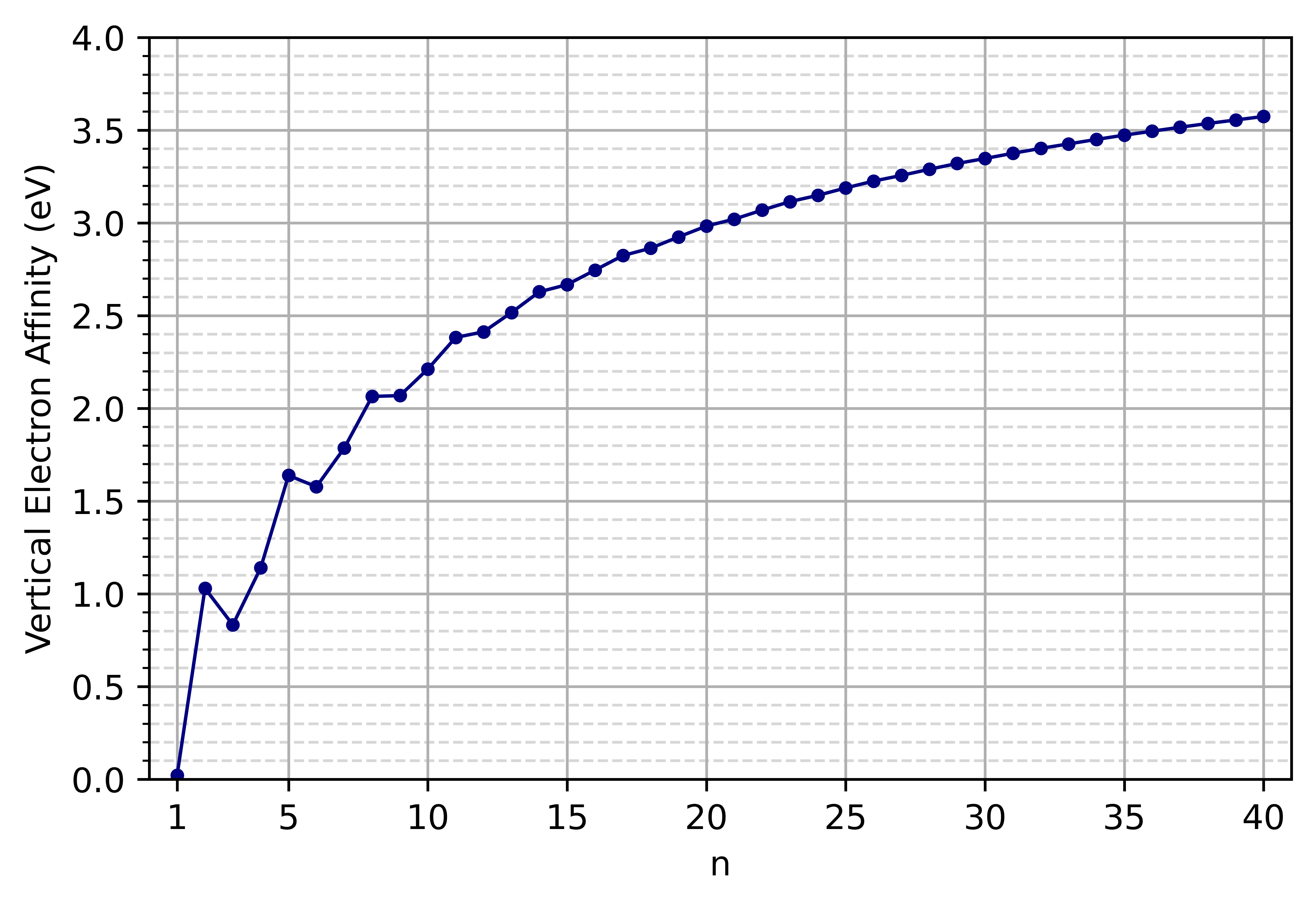} 
\caption{\label{ea} 
Vertical electron affinity of ground-state $n$-LP, calculated using spin-unrestricted TAO-LDA.} 
\end{figure}

\begin{figure} 
\includegraphics[scale=1.0]{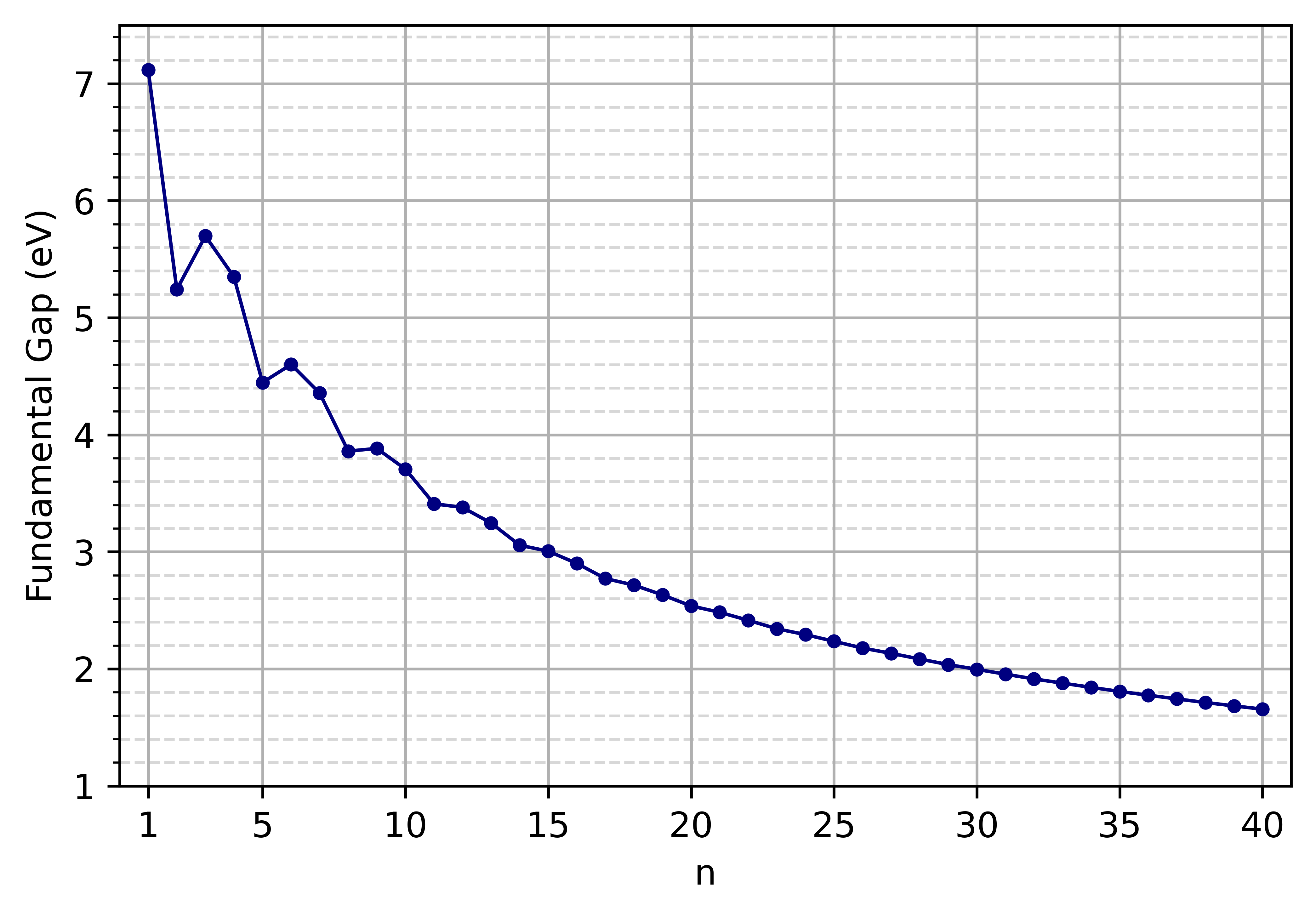} 
\caption{\label{fg} 
Fundamental gap of ground-state $n$-LP, calculated using spin-unrestricted TAO-LDA.} 
\end{figure}

Similar to the trends observed for the ST gaps, the $\text{IP}_{v}$ (see \Cref{ip}) and $\text{EA}_{v}$ (see \Cref{ea}) do not exhibit a simple monotonic evolution with increasing length $n$. Instead, they display distinct 
damped oscillations for shorter $n$-LPs, before converging to an asymptotic limit (also see Table S2 in the SI). Specifically, with increasing $n$-LP length, the $\text{IP}_{v}$ generally decreases, while the 
$\text{EA}_{v}$ generally increases, leading to an overall reduction in $E_{g}$ (see \Cref{fg}). This oscillatory behavior stands in stark contrast to the strictly monotonic trends previously reported for conventional 
nanographenes \citep{ref-TAO2,ref-TAO_GNP}. 

As presented in \Cref{fg}, the $E_{g}$ values of ground-state $n$-LPs ($n$ = 1–40) range from 1.66 eV to 7.12 eV (also see Table S2 in the SI). These electronic properties are critical for determining the suitability of 
electronic systems in optoelectronic applications. Typically, electronic systems with an $E_{g}$ in the 1--3 eV range are considered highly desirable for photovoltaic applications \citep{ref-PV1,ref-PV2}. According to our 
TAO-LDA results, $n$-LPs ($n$ = 16--40) fall within this desirable range, suggesting that the elongated $n$-LPs can be promising candidates for organic solar cells and other nanoscale electronic devices.

\subsection{Symmetrized von Neumann Entropy} 

To quantitatively assess the radical character of ground-state $n$-LP, we calculate the symmetrized von Neumann entropy ($S_{\text{vN}}$) \citep{ref-TAO2,ref-TAO4,ref-TAO_GNP}: 
\begin{equation}\label{S_vN_eq} 
S_{\text{vN}} = - \sum_{i=1}^{\infty} \bigg\lbrace \frac{f_{i}}{2}\ \text{ln}\bigg( \frac{f_{i}}{2} \bigg) + \bigg( 1-\frac{f_{i}}{2} \bigg)\ \text{ln}\bigg( 1-\frac{f_{i}}{2} \bigg) \bigg\rbrace, 
\end{equation} 
where $f_{i}$ (i.e., a value ranging from 0 to 2) is the occupation number of the $i^\text{th}$ orbital, obtained with spin-restricted TAO-LDA. This measure is justified because the TAO-LDA orbital occupation numbers closely 
approximate the NOONs \citep{ref-TAO1,ref-TAO_PAH,ref-NOON}. As shown in Eq.\ (\ref{S_vN_eq}), the occupation numbers ($f_{i}$) of exactly 0 or 2 do not contribute to the $S_{\text{vN}}$, while the $f_{i}$ values that 
strongly deviate from 0 and 2 (e.g., in the range of 0.2–1.8) lead to a significant increase in the $S_{\text{vN}}$. Consequently, for a system with nonradical character (e.g., a stable, closed-shell molecule), all $f_{i}$ values 
remain very close to 0 or 2, resulting in a negligible $S_{\text{vN}}$. By contrast, for a system with pronounced radical character, some $f_{i}$ values can strongly deviate from 0 and 2 (e.g., in the range of 0.2–1.8), and hence, 
the $S_{\text{vN}}$ can substantially increase, as the number of $f_{i}$ values strongly deviating from 0 and 2 grows. Although other quantitative measures of radical character 
exist \citep{ref-Takatsuka,ref-HeadGordon,ref-Yamaguchi,ref-Nakano}, the $S_{\text{vN}}$ (given by Eq.\ (\ref{S_vN_eq})) is particularly useful, since it is closely related to the entropy contribution (i.e., the approximate static 
correlation energy) in TAO-DFT \citep{ref-TAO1,ref-TAO2}, and can be readily obtained without additional computational cost. 

\begin{figure} 
\includegraphics[scale=1.0]{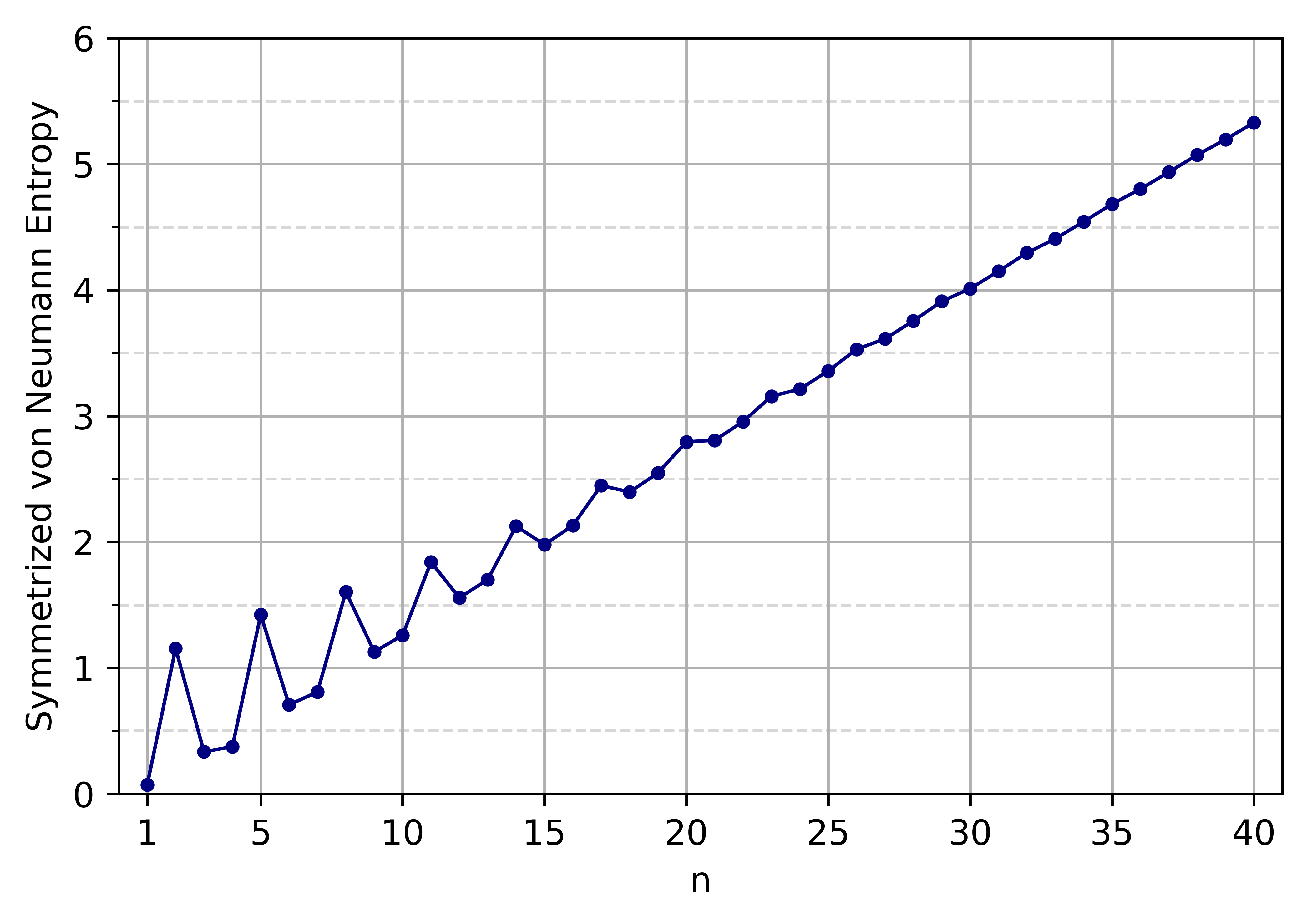} 
\caption{\label{S_vN} 
Symmetrized von Neumann entropy of ground-state $n$-LP, calculated using spin-restricted TAO-LDA.} 
\end{figure}

As presented in \Cref{S_vN}, the $S_{\text{vN}}$ values of ground-state $n$-LPs ($n$ = 1--40) generally increase with increasing length $n$ (also see Table S3 in the SI), suggesting that the radical character of ground-state 
$n$-LPs generally intensifies with chain extension. Consistent with the trends observed in the ST gaps, the $S_{\text{vN}}$ follows a distinct damped oscillatory pattern rather than a purely monotonic increase typically seen in 
conventional nanographenes \citep{ref-TAO2,ref-TAO_GNP}. This periodic behavior is most pronounced for the $S_{\text{vN}}$ values of shorter $n$-LPs, where local maxima are observed at specific lengths of $n = 3k - 1$, 
with $k$ being positive integers. These peaks in $S_{\text{vN}}$ coincide with the sharp dips previously observed in the ST gaps, confirming that the electronic structure at these specific lengths is characterized by a significant 
enhancement of the radical character. As the length $n$ increases, these oscillations gradually attenuate, and the $S_{\text{vN}}$ continues a steady upward trend, reaching 5.33 at $n = 40$.

\subsection{Active Orbital Occupation Numbers} 

To provide further insight into the increase in $S_{\text{vN}}$ with increasing chain length $n$ and to clarify the polyradical character of ground-state $n$-LPs, we examine the active orbital occupation numbers obtained with 
spin-restricted TAO-LDA. As aforementioned, the TAO-LDA orbital occupation numbers serve as the approximate NOONs \citep{ref-TAO1,ref-TAO_PAH,ref-NOON}. For the ground state of $n$-LP (containing $N$ electrons), 
the highest occupied molecular orbital (HOMO) and the lowest unoccupied molecular orbital (LUMO) are defined as the $(N/2)^\text{th}$ and $(N/2+1)^\text{th}$ orbitals, respectively. 

\begin{figure} 
\includegraphics[scale=1.0]{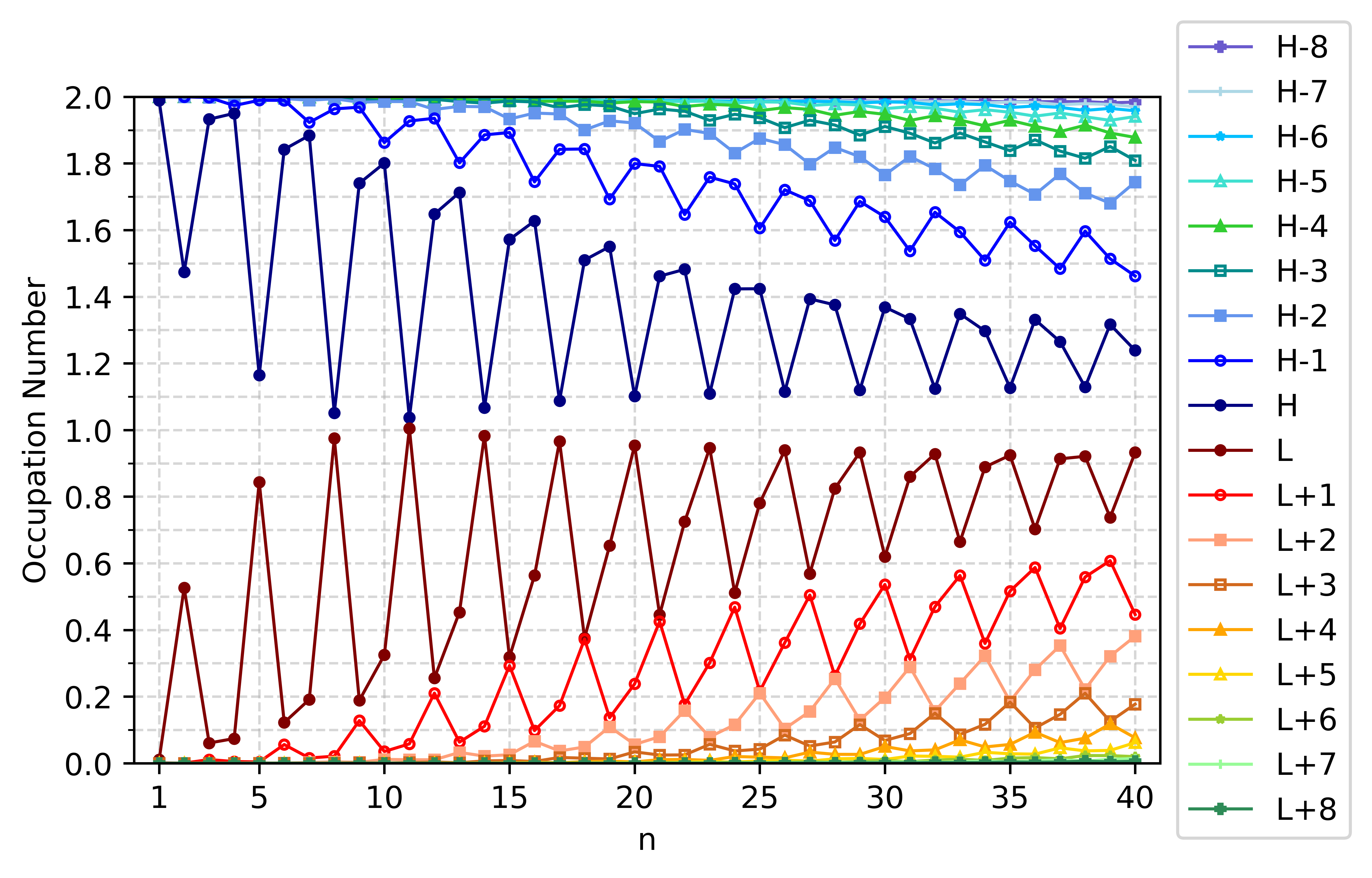} 
\caption{\label{occ} 
Occupation numbers of active orbitals (HOMO$-$8, HOMO$-$7, ..., HOMO, LUMO, ..., LUMO+7, and LUMO+8) for the ground state of $n$-LP, calculated using spin-restricted TAO-LDA. 
For brevity, the HOMO and LUMO are denoted as the H and L, respectively.} 
\end{figure}

As depicted in \Cref{occ}, for the shortest 1-LP, all orbital occupation numbers remain very close to 0 or 2, indicating a closed-shell, nonradical character. This is consistent with the larger $E_{\text{ST}}$ and negligible 
$S_{\text{vN}}$ values associated with the relatively stable 1-LP. However, as the length $n$ increases, the active orbital occupation numbers exhibit a clear damped oscillatory pattern with a periodicity of 3, distinguishing 
$n$-LPs from the monotonic orbital fractionalization typically observed in conventional nanographenes \citep{ref-TAO2,ref-TAO_GNP}. At specific lengths of $n = 3k - 1$ (where $k$ are positive integers), the occupation 
numbers of the HOMO and LUMO show sharp deviations from 2 and 0, respectively, rapidly approaching 1 (half-filling). As the length $n$ increases further, the oscillations gradually attenuate, and the number of orbitals with 
occupation numbers in the range of 0.2--1.8 steadily increases. Similar to the asymptotic behavior of other extended $\pi$-conjugated networks \citep{ref-TAO2,ref-TAO3,ref-TAO4}, this convergence signals a transition from 
a nonradical to an increasingly polyradical character with increasing $n$. This polyradical transition fundamentally accounts for the observed reduction in the $E_{\text{ST}}$ and the steady increase in $S_{\text{vN}}$ for 
longer $n$-LPs.

\subsection{Real-Space Representation of Active Orbitals} 

To spatially observe the evolution of the radical character, we visualize the real-space representation of active orbitals (HOMO$-$4, HOMO$-$3, ..., HOMO, LUMO, ..., LUMO+3, and LUMO+4) for the ground states of some 
representative $n$-LPs (with $n$ = 1--5), obtained with spin-restricted TAO-LDA. 

\begin{figure} 
\includegraphics[scale=0.5]{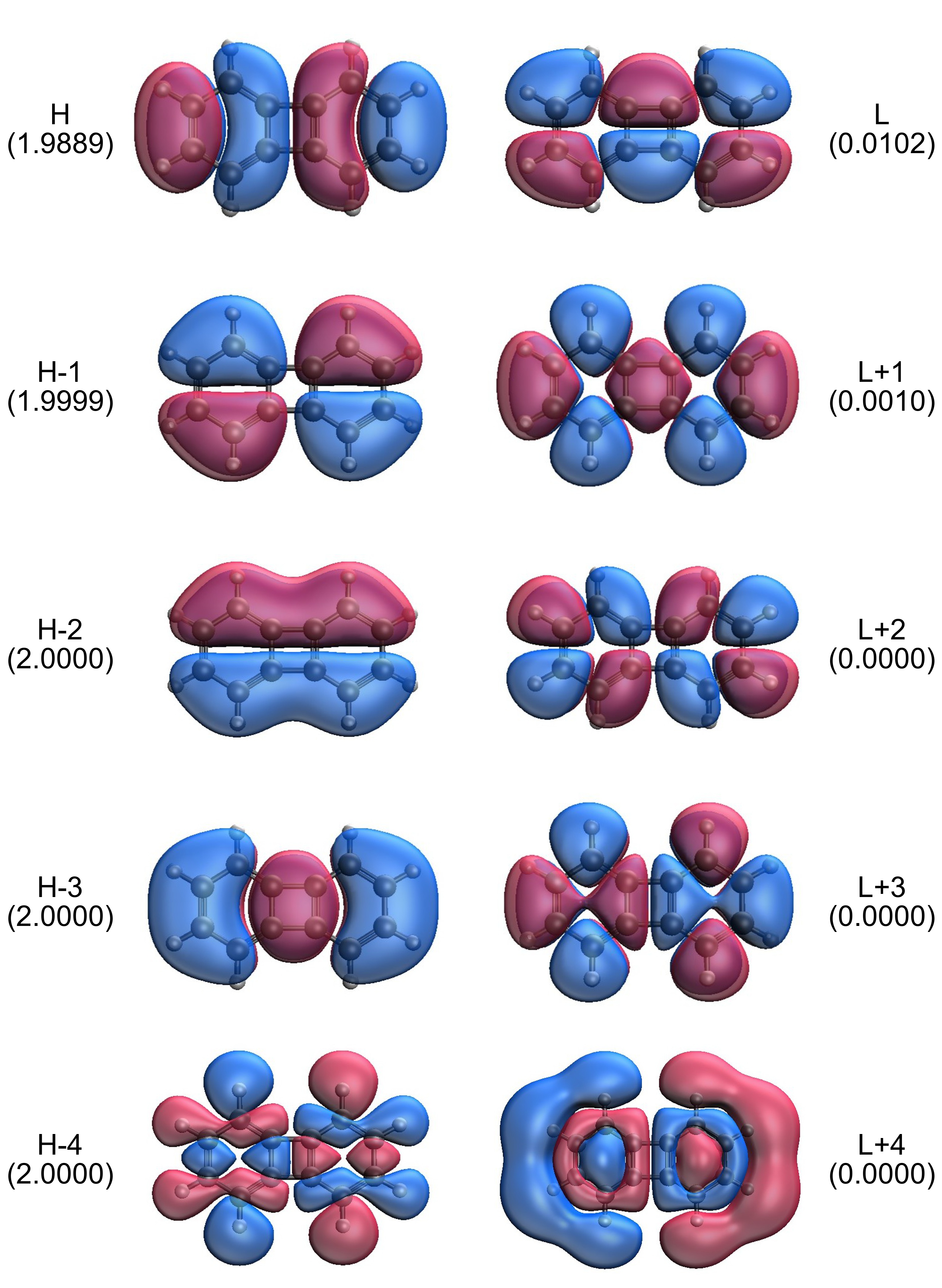} 
\caption{\label{n1_orbital_final} 
Real-space representation of active orbitals (HOMO$-$4, HOMO$-$3, ..., HOMO, LUMO, ..., LUMO+3, and LUMO+4) for the ground state of 1-LP, 
obtained with spin-restricted TAO-LDA, at isovalue = 0.02 $e/\text{\AA}^{3}$, where the orbital occupation numbers are given in parentheses. 
For brevity, the HOMO and LUMO are denoted as the H and L, respectively.} 
\end{figure}

\begin{figure}
\includegraphics[scale=0.5]{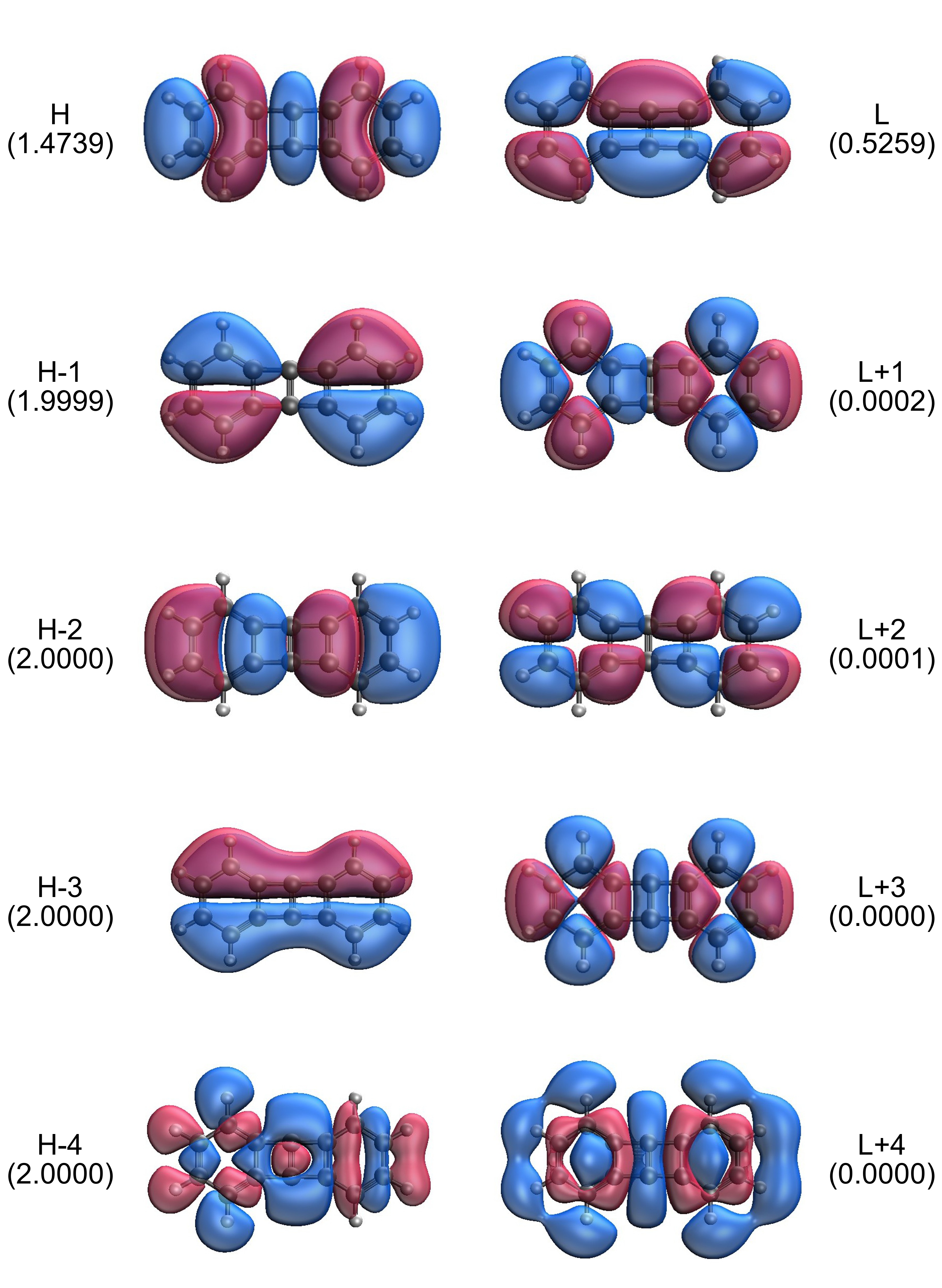} 
\caption{\label{n2_orbital_final} 
Real-space representation of active orbitals (HOMO$-$4, HOMO$-$3, ..., HOMO, LUMO, ..., LUMO+3, and LUMO+4) for the ground state of 2-LP, 
obtained with spin-restricted TAO-LDA, at isovalue = 0.02 $e/\text{\AA}^{3}$, where the orbital occupation numbers are given in parentheses. 
For brevity, the HOMO and LUMO are denoted as the H and L, respectively.} 
\end{figure}

\begin{figure} 
\includegraphics[scale=0.5]{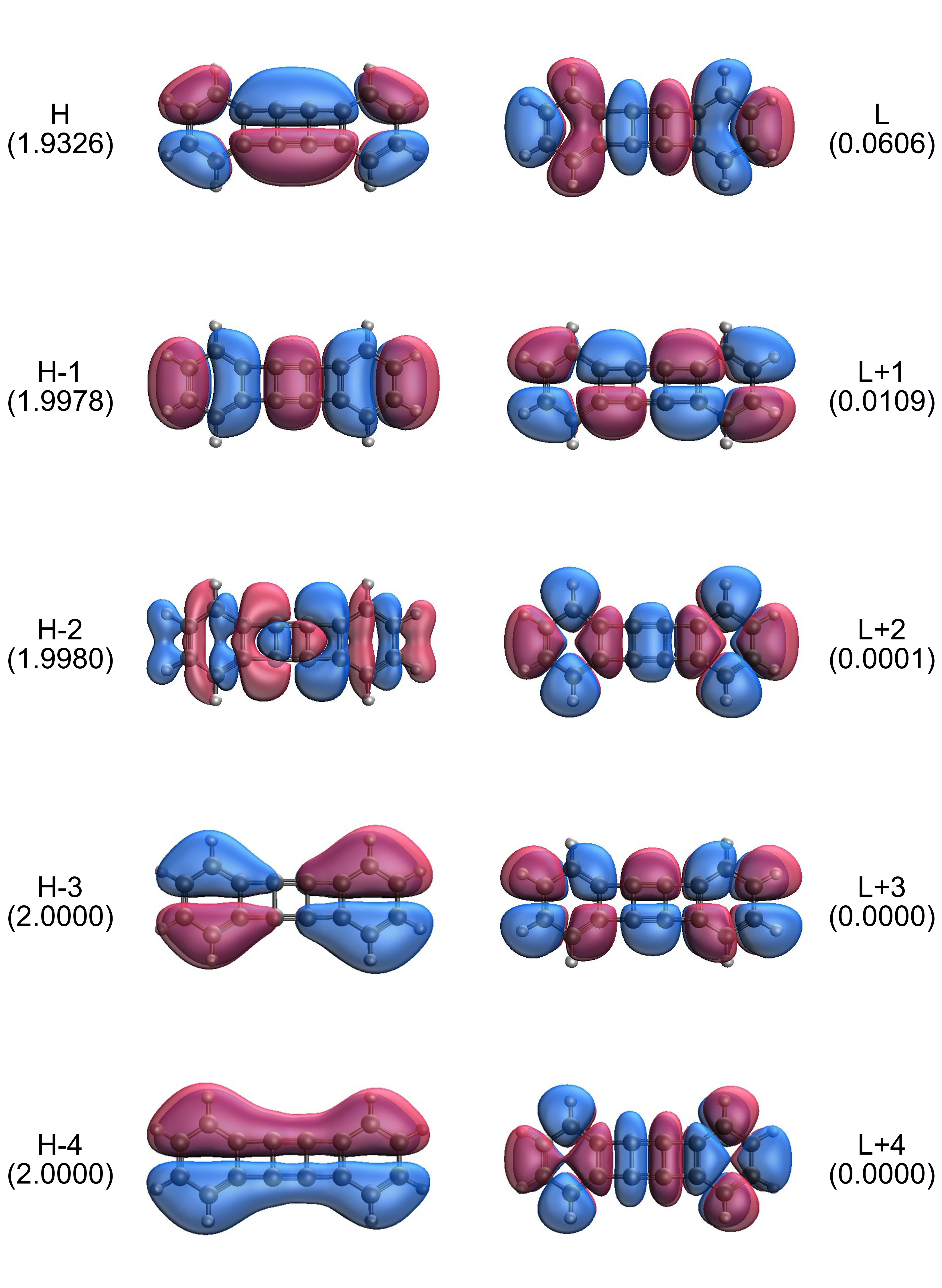} 
\caption{\label{n3_orbital_final} 
Real-space representation of active orbitals (HOMO$-$4, HOMO$-$3, ..., HOMO, LUMO, ..., LUMO+3, and LUMO+4) for the ground state of 3-LP, 
obtained with spin-restricted TAO-LDA, at isovalue = 0.02 $e/\text{\AA}^{3}$, where the orbital occupation numbers are given in parentheses. 
For brevity, the HOMO and LUMO are denoted as the H and L, respectively.} 
\end{figure}

\begin{figure} 
\includegraphics[scale=0.5]{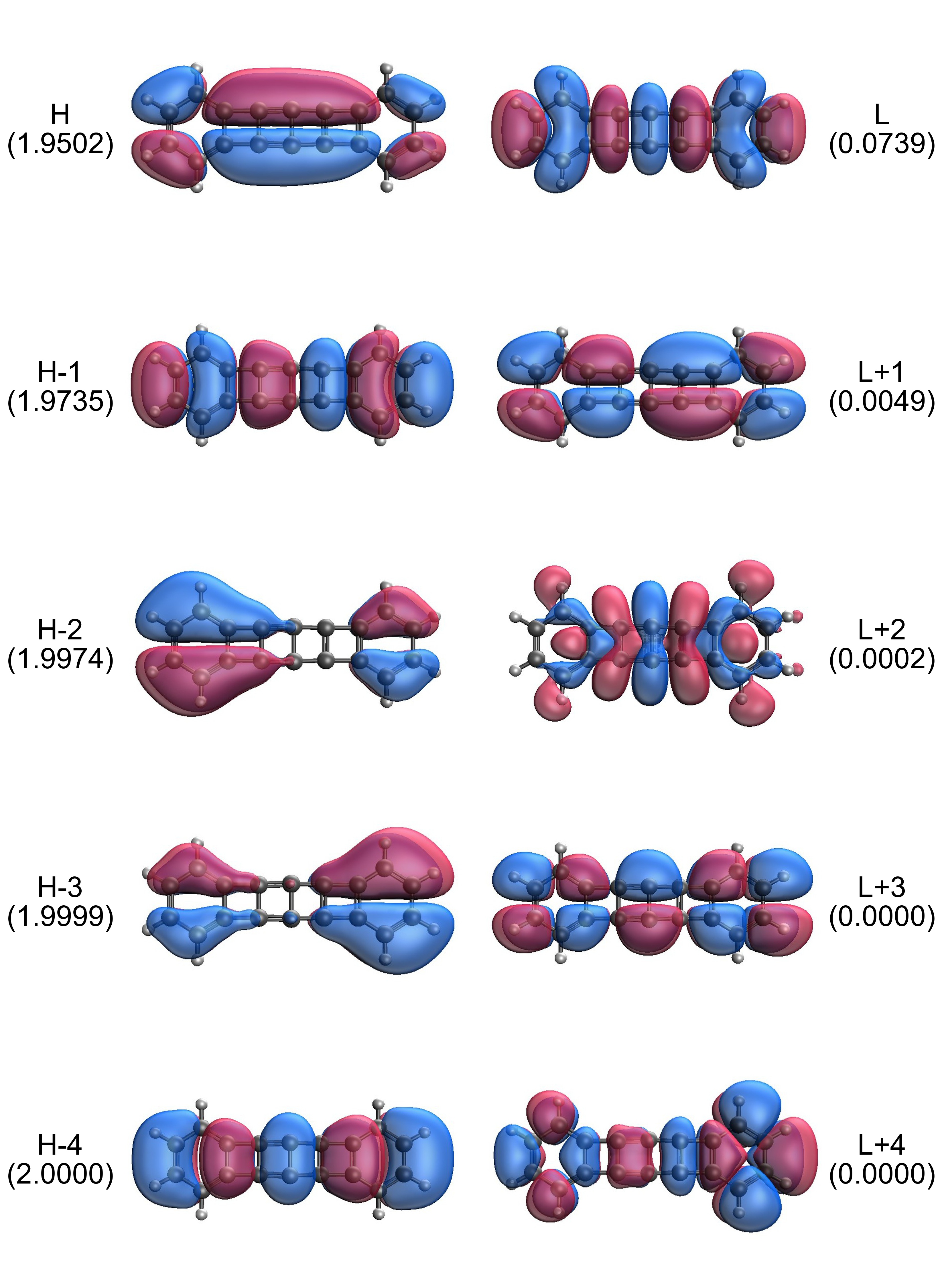} 
\caption{\label{n4_orbital_final} 
Real-space representation of active orbitals (HOMO$-$4, HOMO$-$3, ..., HOMO, LUMO, ..., LUMO+3, and LUMO+4) for the ground state of 4-LP, 
obtained with spin-restricted TAO-LDA, at isovalue = 0.02 $e/\text{\AA}^{3}$, where the orbital occupation numbers are given in parentheses. 
For brevity, the HOMO and LUMO are denoted as the H and L, respectively.} 
\end{figure}

\begin{figure} 
\includegraphics[scale=0.5]{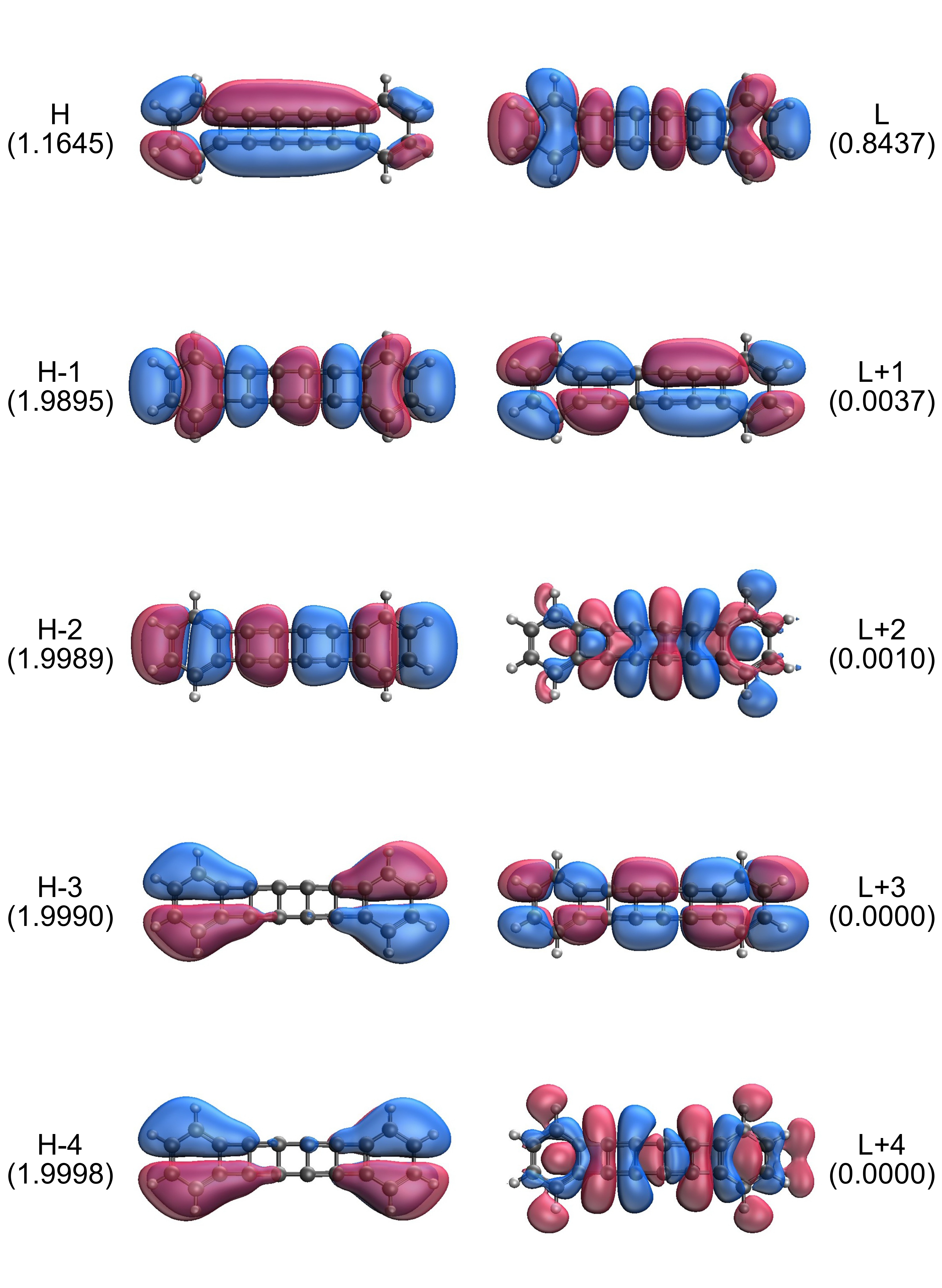} 
\caption{\label{n5_orbital_final} 
Real-space representation of active orbitals (HOMO$-$4, HOMO$-$3, ..., HOMO, LUMO, ..., LUMO+3, and LUMO+4) for the ground state of 5-LP, 
obtained with spin-restricted TAO-LDA, at isovalue = 0.02 $e/\text{\AA}^{3}$, where the orbital occupation numbers are given in parentheses. 
For brevity, the HOMO and LUMO are denoted as the H and L, respectively.} 
\end{figure}

As shown in \Cref{n1_orbital_final,n2_orbital_final,n3_orbital_final,n4_orbital_final,n5_orbital_final}, the spatial distribution of the active orbitals 
reveals a fascinating physical mechanism that governs the electronic properties of $n$-LPs. For the shortest 1-LP, the frontier orbitals (HOMO and LUMO) are fully delocalized over the entire molecular framework. As the 
chain length increases to the first radical hotspot $n$ = 2, the HOMO and LUMO retain their globally delocalized character. Notably, a distinct orbital curve crossing occurs within the deeper occupied levels: the delocalized 
state (HOMO$-$2 in 1-LP) and the state with distinct internal nodes (HOMO$-$3 in 1-LP) swap their energy ordering to become the HOMO$-$3 and HOMO$-$2 states in 2-LP, respectively. 

Despite the absence of spatial localization at the terminal edges, 2-LP exhibits a pronounced polyradical character. This spatial consistency of the active space persists in the subsequent longer chains. As the chain length 
increases to $n$ = 3 and $n$ = 4, the frontier orbitals maintain their delocalized nature, while the system reverts to a predominantly closed-shell, nonradical configuration. At the next radical hotspot $n$ = 5, the strongly 
correlated HOMO and LUMO again remain fully delocalized across the extended central ladder framework, generating significant radical character without edge-localization. 

This visual evidence provides a profound physical insight. Unlike conventional graphene nanosystems (e.g., linear acenes, zigzag GNRs, and GNPs) \citep{ref-TAO2,ref-TAO_GNP}, where polyradical character is typically 
driven by the emergence of edge-localized states near the Fermi level, the radical hotspots ($n = 3k - 1$, with $k$ being positive integers) in $n$-LPs are intrinsically bulk-driven. The periodic emergence of strong static 
correlation is governed by the structural periodicity and the resulting narrowing of the fundamental gap, rather than spatial electron localization at the terminal phenyl rings. This unique, completely delocalized polyradical 
mechanism fundamentally distinguishes $n$-LPs from the monotonic trends observed in traditional nanoribbons. 

Furthermore, a detailed examination of the orbital nodal structures reveals a highly systematic phase evolution that accompanies this chain elongation. First, an alternating spatial parity is observed between consecutive chain 
lengths (e.g., the apparent swapping of HOMO and LUMO character between $n$ and $n+1$), which is a characteristic signature of systematic node addition in extended one-dimensional $\pi$-systems. Second, orbitals 
possessing a longitudinal nodal plane (characterized by an up-down antisymmetric phase distribution) are systematically shifted to deeper occupied energy levels as $n$ increases, specifically from HOMO$-$2 in 1-LP to 
HOMO$-$3 in 2-LP, and subsequently to HOMO$-$4 in 3-LP. This energetic stabilization is driven by the extended longitudinal $\pi$-conjugation, which energetically favors orbitals lacking horizontal nodes. Finally, as the 
central ladder framework extends, the edge-localized states decouple due to spatial separation and manifest as nearly degenerate pairs of symmetric and antisymmetric linear combinations. This phenomenon is vividly 
observed in the deeper levels of longer chains, such as the paired HOMO$-$2/HOMO$-$3 in 4-LP and HOMO$-$3/HOMO$-$4 in 5-LP, where the paired states share identical spatial distributions but feature opposite phase 
combinations across the molecular center. This intricate parity and nodal phase evolution further underscores the complex, size-dependent quantum behavior intrinsic to the $n$-LP structural framework.

\section{Conclusions} 

In conclusion, accurate prediction of the electronic properties of extended $\pi$-conjugated networks with radical character, such as $n$-LPs, presents a significant challenge for conventional electronic structure methods. 
As the length of $n$-LP increases, the radical character becomes more pronounced, rendering standard KS-DFT unreliable, frequently resulting in unphysical spin-symmetry breaking and severe spin contamination in 
spin-unrestricted calculations. Conversely, while high-level {\it ab initio} MR electronic structure methods can capture these static correlation effects, their prohibitive computational cost precludes their applications to 
polyradical nanosystems. To address these limitations, we have adopted TAO-LDA \citep{ref-TAO1} with a nonempirical fictitious temperature ($\theta = 9.55301$ mhartree) \citep{ref-new_theta} to systematically investigate 
the electronic properties of $n$-LPs ($n$ = 1--40). Owing to its decent balance between computational efficiency and accuracy for polyradical nanosystems, TAO-LDA has proved to be a highly suitable method for this study. 

Based on our TAO-LDA calculations, all investigated $n$-LPs ($n$ = 1--40) possess singlet ground states. As the length of $n$-LP increases, the vertical electron affinity and the symmetrized von Neumann entropy generally 
increase, while the singlet-triplet energy gap, the vertical ionization potential, and the fundamental gap generally decrease. Besides, there is a distinct transition from the nonradical character of shorter $n$-LPs to the 
increasing polyradical character of longer $n$-LPs. Notably, in contrast to the strictly monotonic trends observed in many previously studied nanographenes, the electronic properties and radical character of $n$-LPs exhibit 
a unique damped periodic oscillation. We identified highly correlated hotspots that occur at specific lengths $n = 3k - 1$, with $k$ being positive integers. Furthermore, the real-space representation of active orbitals 
demonstrates that this global electronic behavior is governed by the intrinsic structural periodicity and completely delocalized nature of the active orbitals along the central chain of $n$ four-membered rings sandwiched 
between two terminal benzene rings. 

While we have theoretically demonstrated the distinct length-dependent oscillations and polyradical nature of pristine $n$-LPs, the electronic properties of these topological networks can potentially be further engineered. 
Future investigations will explore how chemical doping, edge functionalization, or the introduction of structural defects might influence the periodic resonance matching in these ladder-type nanomaterials.

\section*{Supporting Information Available} 
Detailed TAO-LDA results for the singlet-triplet energy gap ($E_{\text{ST}}$), vertical ionization potential ($\text{IP}_{v}$), vertical electron affinity ($\text{EA}_{v}$), fundamental gap ($E_{g}$), and 
symmetrized von Neumann entropy ($S_{\text{vN}}$) of $n$-LP ($n$ = 1–40) are provided. 

\section*{Data Availability Statement} 
The data that support the findings of this study are available within the article and its Supporting Information. 

\section*{Notes} 
The authors declare no competing financial interest.

\begin{acknowledgments} 

This work was supported by National Science and Technology Council of Taiwan (Grant No.: NSTC115-2112-M-002-011), National Taiwan University, and National Center for Theoretical Sciences of Taiwan. 

\end{acknowledgments}

\newpage 

\newpage 

\newpage 

\newpage 

\newpage 

\newpage 

\newpage 

\newpage 

\newpage 

\newpage 

\newpage 

\newpage 


\begin{thebibliography}{99} 
\expandafter\ifx\csname url\endcsname\relax 
\def\url#1{\texttt{#1}}\fi 
\expandafter\ifx\csname urlprefix\endcsname\relax\def\urlprefix{URL }\fi
\providecommand{\bibinfo}[2]{#2}
\providecommand{\eprint}[2][]{\url{#2}}


\section*{References}

\bibitem[]{ref-carbon1} 
Hirsch, A. 
The era of carbon allotropes. 
{\em Nat. Mater.} {\bf 2010}, {\em 9}, 868--871. 

\bibitem[]{ref-carbon2} 
Georgakilas, V.; Perman, J. A.; Tucek, J.; Zboril, R. 
Broad family of carbon nanoallotropes: Classification, chemistry, and applications of fullerenes, carbon dots, nanotubes, graphene, nanodiamonds, and combined superstructures. 
{\em Chem. Rev.} {\bf 2015}, {\em 115}, 4744--4822. 

\bibitem[]{ref-carbon3} 
Kroto, H. W.; Heath, J. R.; O'Brien, S. C.; Curl, R. F.; Smalley, R. E. 
C60: Buckminsterfullerene. 
{\em Nature} {\bf 1985}, {\em 318}, 162--163. 

\bibitem[]{ref-carbon4} 
Meirzadeh, E.; Evans, A. M.; Rezaee, M.; Milich, M.; Dionne, C. J.; Darlington, T. P.; et al. 
A few-layer covalent network of fullerenes. 
{\em Nature} {\bf 2023}, {\em 613}, 71--76. 

\bibitem[]{ref-carbon5} 
Bethune, D. S.; Klang, C. H.; de Vries, M. S.; Gorman, G.; Savoy, R.; Vazquez, J.; Beyers, R. 
Cobalt-catalysed growth of carbon nanotubes with single-atomic-layer walls. 
{\em Nature} {\bf 1993}, {\em 363}, 605--607. 

\bibitem[]{ref-carbon6} 
Baughman, R. H.; Zakhidov, A. A.; de Heer, W. A. 
Carbon nanotubes—the route toward applications. 
{\em Science} {\bf 2002}, {\em 297}, 787--792. 

\bibitem[]{ref-carbon7} 
Wallace, P. R. 
The band theory of graphite. 
{\em Phys. Rev.} {\bf 1947}, {\em 71}, 622--634. 

\bibitem[]{ref-carbon8} 
Allen, M. J.; Tung, V. C.; Kaner, R. B. 
Honeycomb carbon: A review of graphene. 
{\em Chem. Rev.} {\bf 2010}, {\em 110}, 132--145. 

\bibitem[]{ref-carbon9} 
Novoselov, K. S.; Fal'ko, V. I.; Colombo, L.; Gellert, P. R.; Schwab, M. G.; Kim, K. 
A roadmap for graphene. 
{\em Nature} {\bf 2012}, {\em 490}, 192--200. 

\bibitem[]{ref-graphene1} 
Neto, A. C.; et al. 
The electronic properties of graphene. 
{\em Rev. Mod. Phys.} {\bf 2009}, {\em 81}, 109. 

\bibitem[]{ref-graphene2} 
Madurani, K. A.; Suprapto, S.; Machrita, N. I.; Bahar, S. L.; Illiya, W.; Kurniawan, F. 
Progress in graphene synthesis and its application: History, challenge and the future outlook for research and industry. 
{\em ECS J. Solid State Sci. Technol.} {\bf 2020}, {\em 9}, 093013. 

\bibitem[]{ref-transistors} 
Britnell, L.; Gorbachev, R. V.; Jalil, R.; Belle, B. D.; Schedin, F.; Mishchenko, A.; Georgiou, T.; Katsnelson, M. I.; Eaves, L.; Morozov, S. V.; et al. 
Field-effect tunneling transistor based on vertical graphene heterostructures. 
{\em Science} {\bf 2012}, {\em 335}, 947--950. 

\bibitem[]{ref-GNR1} 
Houtsma, R. K.; de la Rie, J.; Stöhr, M. 
Atomically precise graphene nanoribbons: Interplay of structural and electronic properties. 
{\em Chem. Soc. Rev.} {\bf 2021}, {\em 50}, 6541--6568. 

\bibitem[]{ref-GNR2} 
Barone, V.; Hod, O.; Scuseria, G. E. 
Electronic structure and stability of semiconducting graphene nanoribbons. 
{\em Nano Lett.} {\bf 2006}, {\em 6}, 2748--2754. 

\bibitem[]{ref-TAO2} 
Wu, C.-S.; Chai, J.-D. 
Electronic properties of zigzag graphene nanoribbons studied by TAO-DFT. 
{\em J. Chem. Theory Comput.} {\bf 2015}, {\em 11}, 2003--2011. 

\bibitem[]{ref-GNP1} 
Gu, Y.; Wu, X.; Gopalakrishna, T. Y.; Phan, H.; Wu, J. 
Graphene-like molecules with four zigzag edges. 
{\em Angew. Chem. Int. Ed.} {\bf 2018}, {\em 57}, 6541--6545. 

\bibitem[]{ref-GNP2} 
Sandoval-Salinas, M. E.; Bernabeu-Cabañero, R.; Pérez-Jiménez, A. J.; San-Fabián, E.; Sancho-García, J. C. 
Electronic structure of rhombus-shaped nanographenes: System size evolution from closed- to open-shell ground states. 
{\em Phys. Chem. Chem. Phys.} {\bf 2023}, {\em 25}, 11697--11706. 

\bibitem[]{ref-TAO_GNP} 
Seenithurai, S.; Chai, J.-D. 
Electronic properties of graphene nano-parallelograms: A thermally assisted occupation DFT computational study. 
{\em Molecules} {\bf 2024}, {\em 29}, 349. 

\bibitem[]{ref-GQR1} 
Russo, S.; Oostinga, J. B.; Wehenkel, D.; Heersche, H. B.; Sobhani, S. S.; Vandersypen, L. M.; Morpurgo, A. F. 
Observation of Aharonov-Bohm conductance oscillations in a graphene ring. 
{\em Phys. Rev. B} {\bf 2008}, {\em 77}, 085413. 

\bibitem[]{ref-GQR2} 
Recher, P.; Trauzettel, B.; Rycerz, A.; Blanter, Y. M.; Beenakker, C.; Morpurgo, A. 
Aharonov-Bohm effect and broken valley degeneracy in graphene rings. 
{\em Phys. Rev. B} {\bf 2007}, {\em 76}, 235404. 

\bibitem[]{ref-TAO4} 
Chen, C.-C.; Chai, J.-D. 
Electronic properties of hexagonal graphene quantum rings from TAO-DFT. 
{\em Nanomaterials} {\bf 2022}, {\em 12}, 3943. 

\bibitem[]{ref-nanosystems1} 
Son, Y. W.; Cohen, M. L.; Louie, S. G. 
Energy gaps in graphene nanoribbons. 
{\em Phys. Rev. Lett.} {\bf 2006}, {\em 97}, 216803. 

\bibitem[]{ref-nanosystems2} 
Han, M. Y.; Özyilmaz, B.; Zhang, Y.; Kim, P. 
Energy band-gap engineering of graphene nanoribbons. 
{\em Phys. Rev. Lett.} {\bf 2007}, {\em 98}, 206805. 

\bibitem[]{ref-BP1} 
Fan, Q.; Yan, L.; Tripp, M. W.; Krejčí, O.; Dimosthenous, S.; Kachel, S. R.; Chen, M.; Foster, A. S.; Koert, U.; Liljeroth, P.; Gottfried, J. M. 
Biphenylene network: A nonbenzenoid carbon allotrope. 
{\em Science} {\bf 2021}, {\em 372}, 852--856. 

\bibitem[]{ref-BP2} 
Luo, M.; Zhao, Y.; Zhang, Z. 
Electronic and transport properties of biphenylene carbon nanoribbons. 
{\em Phys. Rev. B} {\bf 2021}, {\em 104}, 115412. 

\bibitem[]{ref-LP1} 
Li, D.-Y.; Qiu, X.; Li, S.-W.; Ren, Y.-T.; Zhu, Y.-C.; Shu, C.-H.; Hou, X.-Y.; Liu, M.; Shi, X.-Q.; Qiu, X.; Liu, P.-N. 
Ladder phenylenes synthesized on Au(111) surface via selective [2+2] cycloaddition. 
{\em J. Am. Chem. Soc.} {\bf 2021}, {\em 143}, 12955--12960. 

\bibitem[]{ref-PV1} 
Scharber, M. C.; Mühlbacher, D.; Koppe, M.; Denk, P.; Waldauf, C.; Heeger, A. J.; Brabec, C. J. 
Design rules for donors in bulk-heterojunction solar cells—Towards 10\% energy-conversion efficiency. 
{\em Adv. Mater.} {\bf 2006}, {\em 18}, 789--794. 

\bibitem[]{ref-PV2} 
Bonaccorso, F.; Sun, Z.; Hasan, T.; Ferrari, A. C. 
Graphene photonics and optoelectronics. 
{\em Nat. Photonics} {\bf 2010}, {\em 4}, 611--622. 

\bibitem[]{ref-Spintronics} 
Bogani, L.; Wernsdorfer, W. 
Molecular spintronics using single-molecule magnets. 
{\em Nat. Mater.} {\bf 2008}, {\em 7}, 179--186. 

\bibitem[]{ref-NLO} 
Nakano, M.; Kishi, R.; Ohta, S.; Takebe, A.; Takahashi, H.; Furukawa, S.; Kubo, T.; Morita, Y.; Nakasuji, K.; Yamaguchi, K.; et al. 
Relationship between third-order nonlinear optical properties and diradical character in open-shell singlet systems. 
{\em Phys. Rev. Lett.} {\bf 2007}, {\em 99}, 033001. 

\bibitem[]{ref-KS1} 
Hohenberg, P.; Kohn, W. 
Inhomogeneous electron gas. 
{\em Phys. Rev.} {\bf 1964}, {\em 136}, B864--B871. 

\bibitem[]{ref-KS2} 
Kohn, W.; Sham, L. J. 
Self-consistent equations including exchange and correlation effects. 
{\em Phys. Rev.} {\bf 1965}, {\em 140}, A1133--A1138. 

\bibitem[]{ref-KS3} 
K\"{u}mmel, S.; Kronik, L. 
Orbital-dependent density functionals: Theory and applications. 
{\em Rev. Mod. Phys.} {\bf 2008}, {\em 80}, 3--60. 

\bibitem[]{ref-SC1} 
Cohen, A. J.; Mori-S\'{a}nchez, P.; Yang, W. 
Insights into current limitations of density functional theory. 
{\em Science} {\bf 2008}, {\em 321}, 792--794. 

\bibitem[]{ref-SC2} 
Cohen, A. J.; Mori-S\'{a}nchez, P.; Yang, W. 
Challenges for density functional theory. 
{\em Chem. Rev.} {\bf 2012}, {\em 112}, 289--320. 

\bibitem[]{ref-Rivero} 
Rivero, P.; Jiménez-Hoyos, C. A.; Scuseria, G. E. 
Entanglement and polyradical character of polycyclic aromatic hydrocarbons predicted by projected Hartree-Fock theory. 
{\em J. Phys. Chem. B} {\bf 2013}, {\em 117}, 12750--12758. 

\bibitem[]{ref-SC3} 
Hachmann, J.; Dorando, J. J.; Avil\'{e}s, M.; Chan, G.K.L. 
The radical character of the acenes: A density matrix renormalization group study. 
{\em J. Chem. Phys.} {\bf 2007}, {\em 127}, 134309. 

\bibitem[]{ref-ab2} 
Mizukami, W.; Kurashige, Y.; Yanai, T. 
More $\pi$ electrons make a difference: Emergence of many radicals on graphene nanoribbons studied by ab initio DMRG theory. 
{\em J. Chem. Theory Comput.} {\bf 2013}, {\em 9}, 401--407. 

\bibitem[]{ref-2RDM1} 
Gidofalvi, G.; Mazziotti, D. A. 
Active-space two-electron reduced-density-matrix method: Complete active-space calculations without diagonalization of the $N$-electron Hamiltonian. 
{\em J. Chem. Phys.} {\bf 2008}, {\em 129}, 134108. 

\bibitem[]{ref-2RDM2} 
Pelzer, K.; Greenman, L.; Gidofalvi, G.; Mazziotti, D. A. 
Strong correlation in acene sheets from the active-space variational two-electron reduced density matrix method: Effects of symmetry and size. 
{\em J. Phys. Chem. A} {\bf 2011}, {\em 115}, 5632--5640. 

\bibitem[]{ref-CAS} 
Roos, B. O.; Taylor, P. R.; Siegbahn, P. E. M. 
A complete active space SCF method (CASSCF) using a density matrix formulated super-CI approach. 
{\em Chem. Phys.} {\bf 1980}, {\em 48}, 157--173. 

\bibitem[]{ref-TAO1} 
Chai, J.-D. 
Density functional theory with fractional orbital occupations. 
{\em J. Chem. Phys.} {\bf 2012}, {\em 136}, 154104. 

\bibitem[]{ref-FTDFT} 
Mermin, N. D. 
Thermal properties of the inhomogeneous electron gas. 
{\em Phys. Rev.} {\bf 1965}, {\em 137}, A1441--A1443. 

\bibitem[]{ref-GGA} 
Chai, J.-D. 
Thermally-assisted-occupation density functional theory with generalized-gradient approximations. 
{\em J. Chem. Phys.} {\bf 2014}, {\em 140}, 18A521. 

\bibitem[]{ref-hybrid1} 
Chai, J.-D. 
Role of exact exchange in thermally-assisted-occupation density functional theory: A proposal of new hybrid schemes. 
{\em J. Chem. Phys.} {\bf 2017}, {\em 146}, 044102. 

\bibitem{Xuan2019} 
Xuan, F.; Chai, J.-D.; Su, H. 
Local density approximation for the short-range exchange free energy functional. 
{\em ACS Omega} {\bf 2019}, {\em 4}, 7675--7683. 

\bibitem{tao-b97-d4} 
Li, S.; Chai, J.-D. 
Optimization of B97-type density functional approximation, global hybrid, and range-separated hybrid energy functionals with the D4 dispersion corrections in TAO-DFT. 
{\em J. Chem. Theory Comput.} {\bf 2025}, {\em 21}, 9538--9572. 

\bibitem[]{ref-new_theta} 
Chen, B.-J.; Chai, J.-D. 
TAO-DFT fictitious temperature made simple. 
{\em RSC Adv.} {\bf 2022}, {\em 12}, 12193--12210. 

\bibitem{etheta-scf} 
Lin, C.-Y.; Hui, K.; Chung, J.-H.; Chai, J.-D. 
Self-consistent determination of the fictitious temperature in thermally-assisted-occupation density functional theory. 
{\em RSC Adv.} {\bf 2017}, {\em 7}, 50496--50507. 

\bibitem[]{ref-TAO_AIMD} 
Li, S.; Chai, J.-D. 
TAO-DFT-based ab initio molecular dynamics. 
{\em Front. Chem.} {\bf 2020}, {\em 8}, 589432. 

\bibitem[]{ref-TAO_PCM} 
Seenithurai, S.; Chai, J.-D. 
TAO-DFT with the polarizable continuum model. 
{\em Nanomaterials} {\bf 2023}, {\em 13}, 1593. 

\bibitem[]{ref-RT_TAO} 
Tsai, H.-Y.; Chai, J.-D. 
Real-time extension of TAO-DFT. 
{\em Molecules} {\bf 2023}, {\em 28}, 7247. 

\bibitem{ft-tao-dft} 
Li, S.; Chai, J.-D. 
Finite-temperature thermally-assisted-occupation density-functional theory, ab initio molecular dynamics, and quantum mechanics and molecular mechanics methods. 
{\em Phys. Rev. A} {\bf 2026}, {\em 113}, 032805. 

\bibitem[]{ref-TAO_PAH} 
Yeh, C.-N.; Chai, J.-D. 
Role of Kekul\'{e} and non-Kekul\'{e} structures in the radical character of alternant polycyclic aromatic hydrocarbons: A TAO-DFT study. 
{\em Sci. Rep.} {\bf 2016}, {\em 6}, 30562. 

\bibitem[]{ref-NOON} 
L\"{o}wdin, P. O.; Shull, H. 
Natural orbitals in the quantum theory of two-electron systems. 
{\em Phys. Rev.} {\bf 1956}, {\em 101}, 1730--1739. 

\bibitem{tao-symmetry} 
Wang, Y.-Y.; Chai, J.-D. 
Spin symmetry in thermally-assisted-occupation density-functional theory. 
{\em Phys. Rev. A} {\bf 2024}, {\em 109}, 062808. 

\bibitem[]{ref-TAO_acene} 
Seenithurai, S.; Chai, J.-D. 
Effect of Li adsorption on the electronic and hydrogen storage properties of acenes: A dispersion-corrected TAO-DFT study. 
{\em Sci. Rep.} {\bf 2016}, {\em 6}, 33081. 

\bibitem[]{ref-TAO3} 
Wu, C.-S.; Lee, P.-Y.; Chai, J.-D. 
Electronic properties of cyclacenes from TAO-DFT. 
{\em Sci. Rep.} {\bf 2016}, {\em 6}, 37249. 

\bibitem{Manassir2020} 
Manassir, M.; Pakiari, A. H. 
Total non-lewis structures: an application to predict the stability and reactivity of linear and angular polyacenes. 
{\em J. Mol. Graphics Modell.} {\bf 2020}, {\em 99}, 107643. 

\bibitem{tao-vib1} 
Hanson-Heine, M. W. D. 
Static correlation in vibrational frequencies studied using thermally-assisted-occupation density functional theory. 
{\em Chem. Phys. Lett.} {\bf 2020}, {\em 739}, 137012. 

\bibitem{HansonHeine2020a} 
Hanson-Heine, M. W. D.; Rogers, D. M.; Woodward, S.; Hirst, J. D. 
Dewar benzenoids discovered in carbon nanobelts. 
{\em J. Phys. Chem. Lett.} {\bf 2020}, {\em 11}, 3769--3772. 

\bibitem{HansonHeine2020b} 
Hanson-Heine, M. W. D.; Hirst, J. D. 
M\"obius and H\"uckel cyclacenes with Dewar and Ladenburg defects. 
{\em J. Phys. Chem. A} {\bf 2020}, {\em 124}, 5408--5414. 

\bibitem{Bettinger2021} 
T\"onshoff, C.; Bettinger, H. F. 
Pushing the Limits of acene chemistry: The recent surge of large acenes. 
{\em Chem. Eur. J.} {\bf 2021}, {\em 27}, 3193--3212. 

\bibitem{Bettinger2021a} 
Gupta, D.; Omont, A.; Bettinger, H. F. 
Energetics of formation of cyclacenes from 2,3-didehydroacenes and implications for astrochemistry. 
{\em Chem. Eur. J.} {\bf 2021}, {\em 27}, 4605--4616. 

\bibitem{tao-vib2} 
Hanson-Heine, M. W. D. 
Static electron correlation in anharmonic molecular vibrations: A hybrid TAO-DFT study. 
{\em J. Phys. Chem. A} {\bf 2022}, {\em 126}, 7273--7282. 

\bibitem{Lischka2023} 
Nieman, R.; Carvalho, J. R.; Jayee, B.; Hansen, A.; Aquino, A. J.; Kertesz, M.; Lischka, H. 
Polyradical character assessment using multireference calculations and comparison with density-functional derived fractional occupation number weighted density analysis. 
{\em Phys. Chem. Chem. Phys.} {\bf 2023}, {\em 25}, 27380--27393. 

\bibitem{Somani2024} 
Somani, A.; Gupta, D.; Bettinger, H. F. 
Computational studies of dimerization of [$n$]-cyclacenes. 
{\em J. Phys. Chem. A} {\bf 2024}, {\em 128}, 6847--6852. 

\bibitem{tao-vib3} 
Hanson-Heine, M. W. D. 
Vibrational stabilization in cyclacene carbon nanobelts. 
{\em J. Phys. Chem. A} {\bf 2025}, {\em 129}, 8601--8612. 

\bibitem{Somani2025} 
Somani, A.; Gupta, D.; Bettinger, H. F. 
Remarkably efficient [4+4] dimerization of [n]-cyclacenes. 
{\em Chemistry} {\bf 2025}, {\em 7}, 62. 

\bibitem{Somani2026} 
Somani, A.; Gupta, D.; Grunenberg, J.; Bettinger, H. F. 
Cyclacene stability: The interplay of strain, aromaticity and force coupling. 
{\em J. Comput. Chem.} {\bf 2026}, {\em 47}, e70325. 

\bibitem[]{ref-qchem} 
Epifanovsky, E.; Gilbert, A. T. B.; Feng, X.; Lee, J.; Mao, Y.; Mardirossian, N.; Pokhilko, P.; White, A. F.; Coons, M. P.; Dempwolff, A. L.; et al. 
Software for the frontiers of quantum chemistry: An overview of developments in the Q-Chem 5 package. 
{\em J. Chem. Phys.} {\bf 2021}, {\em 155}, 084801. 

\bibitem[]{ref-IQmol} 
Gilbert, A. T. B. 
IQmol: A free open-source molecular builder and visualization package. 
Available online: http://iqmol.org (accessed on 21 May 2026). 

\bibitem[]{ref-ST_device1} 
Bendikov, M.; Duong, H. M.; Starkey, K.; Houk, K. N.; Carter, E. A.; Wudl, F. 
Oligoacenes: Theoretical prediction of open-shell singlet diradical ground states. 
{\em J. Am. Chem. Soc.} {\bf 2004}, {\em 126}, 7416--7417. 

\bibitem[]{ref-ST_device2} 
Xia, J.; Sanders, S. N.; Cheng, W.; Low, J. Z.; Liu, J.; Campos, L. M.; Sun, T. 
Singlet fission: Progress and prospects in solar cells. 
{\em Adv. Mater.} {\bf 2017}, {\em 29}, 1601652. 

\bibitem[]{ref-ST1} 
Lambert, C. J. 
Basic concepts of quantum interference in molecular electronics. 
{\em Chem. Soc. Rev.} {\bf 2015}, {\em 44}, 875--888. 

\bibitem[]{ref-ST2} 
Poater, J.; Garcia-Cruz, I.; Illas, F.; Bofill, J. M. 
Clar's aromatic sextet theory revisited. 
{\em J. Org. Chem.} {\bf 2005}, {\em 70}, 8065--8070. 

\bibitem[]{ref-Deleuze2003} 
Deleuze, M. S.; Claes, L.; Kryachko, E. S.; Fran\c{c}ois, J. P. 
Benchmark theoretical study of the ionization threshold of benzene and oligoacenes. 
{\em J. Chem. Phys.} {\bf 2003}, {\em 119}, 3106--3119. 

\bibitem[]{ref-Hajgato2008} 
Hajgat\'{o}, B.; Deleuze, M. S.; Tozer, D. J.; De Proft, F. 
A benchmark theoretical study of the electron affinities of benzene and linear acenes. 
{\em J. Chem. Phys.} {\bf 2008}, {\em 129}, 084308. 

\bibitem[]{ref-Takatsuka} 
Takatsuka, K.; Fueno, T.; Yamaguchi, K. 
Distribution of odd electrons in ground state molecules. 
{\em Theor. Chim. Acta} {\bf 1978}, {\em 48}, 175--183. 

\bibitem[]{ref-HeadGordon} 
Head-Gordon, M. 
Characterizing unpaired electrons from the one-particle density matrix. 
{\em Chem. Phys. Lett.} {\bf 2003}, {\em 372}, 508--511. 

\bibitem[]{ref-Yamaguchi} 
Yamaguchi, K. 
The diradical character of the trimethylenemethane model. 
{\em Chem. Phys. Lett.} {\bf 1975}, {\em 33}, 330--335. 

\bibitem[]{ref-Nakano} 
Nakano, M.; Kishi, R.; Nitta, T.; Kubo, T.; Nakasuji, K.; Kamada, K.; Ohta, K.; Champion, R.; Morita, Y.; Yamaguchi, K. 
Diradical character and second hyperpolarizabilities of two-dimensional spatial dictates: Open-shell singlet polycyclic aromatic hydrocarbons. 
{\em J. Phys. Chem. A} {\bf 2005}, {\em 109}, 885--891. 

\end{thebibliography}
\end{document}